\documentclass[aps,preprintnumbers,superscriptaddress,twocolumn,amsmath,amssymb,floatfix,prl]{revtex4}
\pdfoutput=1
\usepackage{graphicx}
\usepackage[usenames,dvipsnames]{color}
\usepackage[colorlinks,bookmarks=false,citecolor=NavyBlue,linkcolor=Red,urlcolor=blue]{hyperref}

\usepackage[export]{adjustbox}

\newcommand\be{\begin{equation}}
\newcommand\bea{\begin{eqnarray}}
\newcommand\bes{\begin{subequations}}
\newcommand\esu{\end{subequations}}
\newcommand\ee{\end{equation}}
\newcommand\eea{\end{eqnarray}}

\newcommand{\cmmnt}[1]{}

\newcommand{\dd}{\text{d}}

\def\red{\color{red}}
\def\doi{http://dx.doi.org/}

\usepackage{braket}

\newcommand\titleinfo{
Generalized hydrodynamics with space-time inhomogeneous interactions
}

\begin{document}

\title{\titleinfo}

\author{Alvise Bastianello}
\affiliation{Institute for Theoretical Physics, University of Amsterdam, Science Park 904, 1098 XH Amsterdam, The Netherlands}

\author{Vincenzo Alba}
\affiliation{Institute for Theoretical Physics, University of Amsterdam, Science Park 904, 1098 XH Amsterdam, The Netherlands}

\author{Jean-S\'ebastien Caux}
\affiliation{Institute for Theoretical Physics, University of Amsterdam, Science Park 904, 1098 XH Amsterdam, The Netherlands}


\begin{abstract}
We provide a new hydrodynamic framework to describe out-of-equilibrium integrable systems with
space-time inhomogeneous interactions. Our result builds up on the recently-introduced Generalized
Hydrodynamics (GHD). The method allows to analytically describe the dynamics during generic
space-time-dependent  smooth  modulations of the interactions.
As a proof of concept, we study experimentally-motivated interaction quenches in
the trapped interacting Bose gas, which cannot be treated with current analytical
or numerical methods. We also benchmark our results in the XXZ spin chain and
in the classical sinh-Gordon model.
\end{abstract}


\maketitle

\paragraph{Introduction. ---}Exploring the out-of-equilibrium behavior of quantum many-body systems  is nowadays among the most active research areas in physics, due
to a successful synergy between theoretical and experimental
advances  \cite{Bloch_rev,Trot12,Lan13,Mein13}.

How, and in what sense, does a coarse-grained thermodynamic description emerge through
dynamical evolution in isolated out-of-equilibrium many-body systems?
One-dimensional systems represent an ideal playground to address this question: there, remarkably powerful
tools exist, both theoretical (such as conformal field theory \cite{difrancesco}
and integrability~\cite{Korepin,smirnov}) and computational (such as Matrix Product States methods~\cite{Sch05}).
Integrability is ubiquitous in the low-dimensional world (and experimentally realized \cite{Giam}),
with applications ranging from spin
chains \cite{Korepin} to continuum models, the latter having Lorentz \cite{smirnov} or Galilean \cite{LiLi63,LiLi63_bis} invariance, or neither \cite{Po01}. 
Integrable models are characterized by the presence of infinitely many conserved charges $\hat{\mathcal{Q}}_j$,  which can be used to exactly determine their thermodynamics \cite{taka}. In recent times, the importance of quasi-local charges has moreover been underlined \cite{IlMePrZa16}.

The last decade has witnessed exact results reaching out-of-equilibrium protocols as well: great attention has been devoted to
the homogeneous sudden quantum quench \cite{calabrese-cardy} (see also Ref. \cite{special_issue} and reference therein).
Because of the conserved quantities, the system exhibits local relaxation to a state that is not thermal \cite{PozMeWe14,MePoz15,DeNaWo14,WoDeNa14,Poz14,PiPoVe17,Po18,PoPiVe18,SoTaMu15}, but rather emerges from a Quench Action \cite{CaEs13,Ca16} or (where applicable) a Generalized Gibbs Ensemble (GGE) \cite{Rigol07,Rigol09} which accounts for all the relevant charges.

More recently, the focus has been on quenches from spatially inhomogeneous systems.
A new theoretical toolbox, dubbed Generalized Hydrodynamics (GHD)~\cite{transportbertini,hydrodoyon1}
allows us to address this problem.
In Refs. \cite{transportbertini,hydrodoyon1} GHD dealt with inhomogeneous states evolving under
a homogeneous Hamiltonian. Several applications have
been explored \cite{GHD3,GHD6,GHD7,GHD8,GHD10,F17,DS,ID117,DDKY17,
DSY17,ID217,CDV17,Bas_Deluca_defhop,Bas_Deluca_defising,PeGa17,CDDK17,
mazza2018,BePC18,Bas2018,Kormos2018,DoyonSphon17,Doyon17,BasDeLu18,BiCoRoDeLuMa19,BePiKo19,DeLuCoDeNa17}, for instance, including diffusive
corrections~\cite{DeBD18,vasseurdiff,GoVa18,AgGoVa19} or applying it to
classical models~\cite{BDWY17,Sp19,Do19,BuCaMo19,CaoBuMo18}. 
Also, combined with the quasiparticle picture for 
integrable systems~\cite{AlCal17,AlCal17a}, GHD allows to 
describe the entanglement spreading after inhomogeneous quenches 
\cite{Al18,BeFAPiCa18,Al19,AlBeFa18}. 
Very recently, it has been shown that GHD provides the correct
theoretical framework to describe atom-chip experiments~\cite{SCBoDoDu19}.

When comparing with actual experiments, inhomogeneities, for instance due to external trapping potentials, 
should ideally be kept into account.
Strictly speaking, inhomogeneities break integrability, but smooth variations 
can still be captured by invoking local relaxation to a (locally homogeneous) integrable model.

Inhomogeneities in the dynamics have already been studied with some limitations for either spatial \cite{GHD3,BuCaMo19} or temporal changes \cite{BasDeLu18}, opening the possibility, for example, of studying the famous Quantum Newton Cradle experiment \cite{exp2} through GHD \cite{CDDK17}.
\begin{figure}[b!]
\includegraphics[width=0.8\columnwidth]{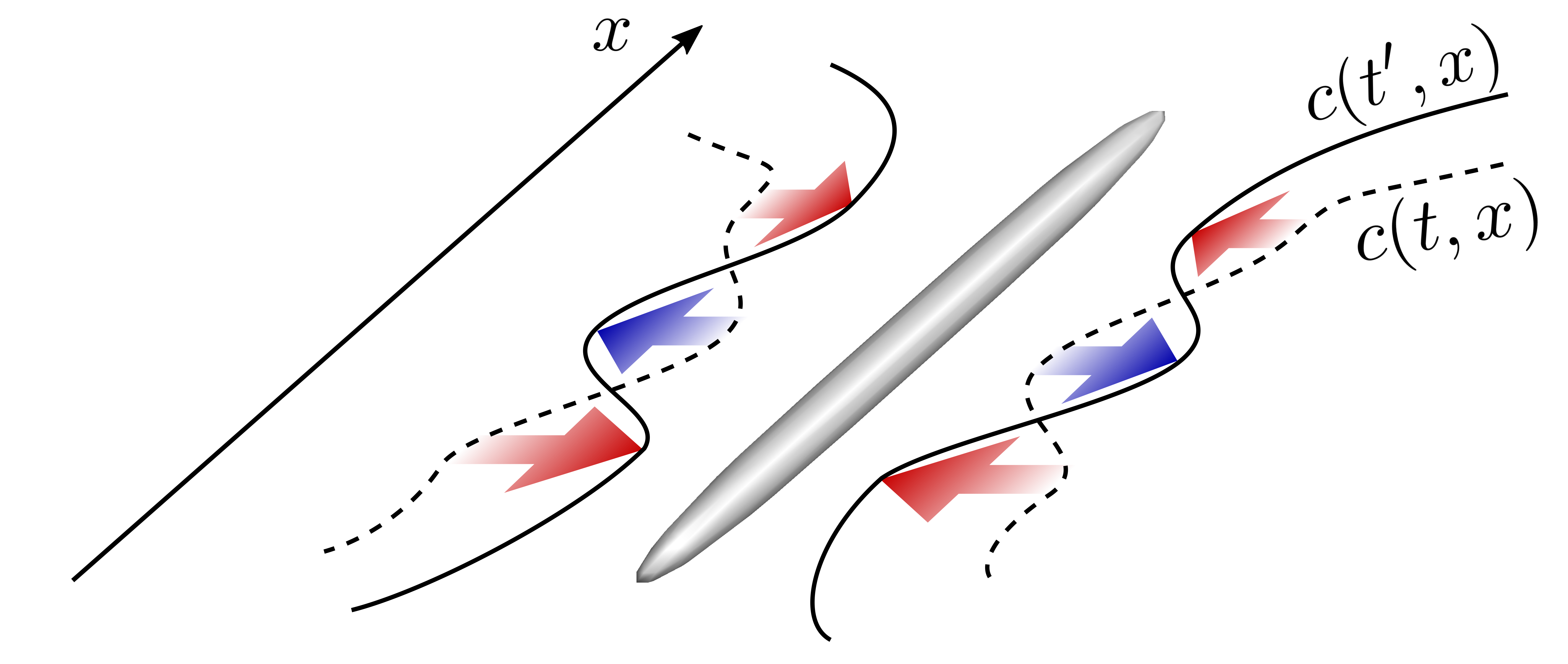}
\caption{\label{fig_cartoon}
Prototypical experimental setup that can be addressed with our method. A Bose gas is trapped in
a one-dimensional tube. The space-time dependent interparticle interaction strength $c(t,x)$ is
modified by modulating the transverse trapping potential (see also Fig. \ref{fig_LL}).
}
\end{figure}
\begin{figure*}
\includegraphics[width=1\textwidth]{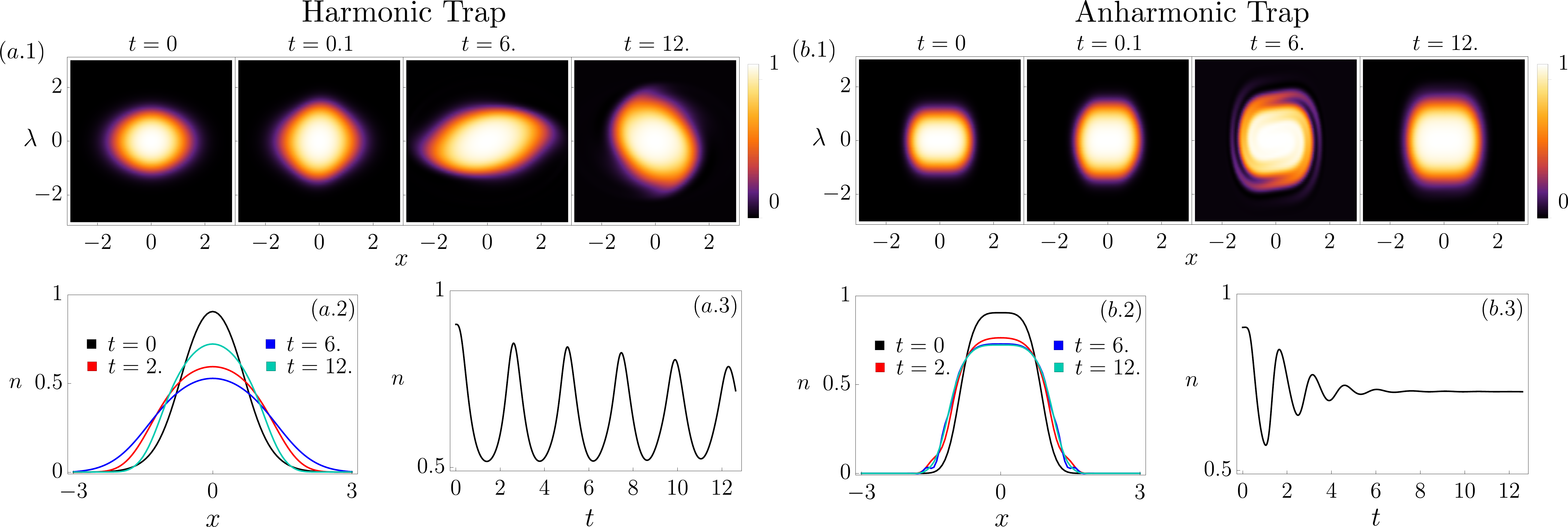}
\caption{\label{fig_LL}
Evolution of the trapped one-dimensional  Lieb-Liniger gas. The interaction strength is changed as $c(t)=0.3+\tanh(3t)$ during
the evolution. The left and right panels correspond to the harmonic and anharmonic trapping potentials $V(x)=x^2/2-0.5$ and $V(x)=x^4/2-0.5$,
respectively. In both cases the initial state is a thermal state at inverse temperature $\beta=2$.
In $(a.1)$ and $(b.1)$  we show the space dependence of the quasiparticle filling functions at different times. In each subfigure, the $y$-axis $\lambda$ is the
quasiparticle rapidity. The $x$-axis shows the position inside the trap.
$(a.2)$ Particle densities $n(t,x)\equiv\langle\psi^\dagger(x)\psi(x)\rangle$ as a function of $x$, for
several times.
Subfig. $(a.3)$: density at the center of the trap as a function of time.
Subfigs. $(b.2)$ and $(b.3)$: the same as in $(a.2)$ and $(a.3)$ for the quench in the anharmonic trap. 
}
\end{figure*}
However, the current state of the art cannot capture changes in the interparticle 
interactions, leaving perturbative methods \cite{EckKo10,MoKe10} or bosonization techniques \cite{PoKo11,DzTy11,ChSc16,DoHaZa11,BeCiKoOr14,PolGR08,BiDeLuViRoMAFA16} as the only methods to tackle this experimentally relevant situation.

In this Letter, we present a complete GHD approach that allows to treat the
dynamics under integrable Hamiltonians with space-time inhomogeneous interactions.
Our results significantly extend the current GHD framework  exhausting all
the possible inhomogeneities which can be considered on a pure hydrodynamic level, disclosing  the full power of GHD in describing experimentally relevant protocols.
We discuss the potential applications of our result to interaction changes in
the Lieb-Liniger model \cite{LiLi63,LiLi63_bis} (Fig.
\ref{fig_cartoon}-\ref{fig_LL}), which is of primary experimental interest.
So far, the primary analytical tool used in dealing with time-dependent interactions has been the Luttinger Liquid approach \cite{Ha81,Ha81_bis}, recently generalized to include spatial inhomogeneities \cite{DuStViCa17,DuStCa17,BrDu18,BrDu17,EiBa17,RuBrDu19,MuRuCa19}. In contrast with GHD, this method is nevertheless confined to the low-energy excitations. 
We numerically benchmark the GHD predictions both in the quantum and classical realms, 
considering the XXZ spin chain and the classical sinh-Gordon field theory, 
showing once again the wide applicability of our results.
Furthermore, we improve the numerical method proposed in Ref. \cite{GHD7} to solve GHD equations, promoting it from a first order to a second order algorithm in the time step, providing a great stability enhancement.

\paragraph{Thermodynamics of integrable models. ---}

The Thermodynamic Bethe Ansatz (TBA) technique is nowadays a textbook topic \cite{taka}: here we present the basic concepts for the sake of a self-contained exposition.
The Hilbert space of integrable models can be understood in terms of multiparticle states  $|\{\lambda\}_{i=1}^N\rangle$, 
labeled by suitable parameters $\lambda$ called rapidities \cite{Korepin,smirnov}. 
Quasiparticles undergo pairwise elastic scatterings, which are described by an interaction-dependent  scattering matrix $S(\lambda)$.
These states are common eigenstates of the full set of \mbox{(quasi-)local} charges. 
In the thermodynamic limit (TDL), we switch to a coarse-grained description through a rapidity (root) density $\rho(\lambda)$ \cite{taka}, which gives the density of rapidities within the interval $(\lambda,\lambda+\dd\lambda)$.
The root densities are in a one-to-one correspondence with the possible thermodynamic states of the system, such as GGEs \cite{IlQuNaBr16} or thermal states and fix the (extensive part of) the expectation value of the local charges
\be\label{eq_ch}
\lim_{\text{TL}}\frac{1}{L}\langle \{\lambda\}_{i=1}^N|\hat{\mathcal{Q}}_j|\{\lambda\}_{i=1}^N\rangle=\int \dd \lambda \, q_j(\lambda) \rho(\lambda)\, ,
\ee
together with any other local (in real space) property of the system, according to the Quench Action approach \cite{CaEs13,Ca16,CuPa19}. The function $q_j(\lambda)$ in Eq. \eqref{eq_ch} is called the charge eigenvalue.

Nontrivial interactions induce collective effects.
For example, the group velocity of the quasiparticles, which is  defined as $v(\lambda)=\partial_\lambda \epsilon/\partial_\lambda p$, 
with $\epsilon(\lambda)$, $p(\lambda)$ the energy and momentum eigenvalues respectively, is 
``dressed'' as $v^\text{eff}(\lambda)=(\partial_\lambda \epsilon)^\text{dr}/(\partial_\lambda p)^\text{dr}$, 
with $\epsilon^{\text{dr}}$ and $p^{\text{dr}}$ the dressed quasiparticle energy and momentum. 
These are obtained by using that an arbitrary dressed quantity $\tau^{\text{dr}}(\lambda)$ is defined through the integral 
equation 
\be
\tau^\text{dr}(\lambda)=\tau(\lambda)-\int \frac{\dd \mu}{2\pi} \partial_\lambda \Theta(\lambda-\mu)\vartheta(\mu)\tau^\text{dr}(\mu)\, ,
\label{dress}
\ee
with $\tau(\lambda)$ the ``bare'' quantity. 
Here $\Theta(\lambda)=-i\log S(\lambda)$, with $S(\lambda)$ the two-body scattering 
matrix encoding the interaction, while $\vartheta(\lambda)=2\pi \rho(\lambda)/(\partial_\lambda p)^{\text{dr}}$ is the so-called filling function.
We summarized the TBA considering a single particle species, but the construction is easily generalized to several species of excitations and bound states.

\paragraph{Emergent hydrodynamics with space-time inhomogeneous interactions. ---}
TBA describes homogeneous stationary states. Instead, we now consider smooth space-time 
inhomogeneities, both in the initial state and in the Hamiltonian. We imagine a family 
of integrable models parametrized by a coupling $\alpha$, 
with Hamiltonians 
\begin{equation}
\hat{H}(\alpha)=\int \dd x\, 
\hat{\bf h}(x,\alpha(t,x)) \, .
\label{eq_inh_H}
\end{equation}
Crucially, in Eq.~\eqref{eq_inh_H} $\alpha$ is a function of both space and time. 
We consider models in the continuum 
for simplicity, but the same construction can be repeated on the lattice.

Spatial inhomogeneities of the initial state on the same typical length-scale of 
the variation of $\alpha$ are allowed. We are then interested in describing the system 
at the Eulerian scales $(\Delta t,\Delta x)\sim ((\partial_t\alpha)^{-1},(\partial_x\alpha)^{-1})$, 
considering at the same time the limit of infinitely smooth variations 
$\partial_t\alpha\sim \partial_x\alpha\to 0$.

Closely following the same argument presented in Refs.~\cite{transportbertini,hydrodoyon1}, 
in this limit we can invoke local relaxation to an inhomogeneous GGE, associated with a 
weakly inhomogeneous root density $\rho(t,x,\lambda)$.
We report the details of the derivation of the GHD equations in the 
Supplemental Material (SM) \cite{suppl}. Here, we rather present the result, discussing 
its physical interpretation and validity regime, together with possible applications.

Our main result is that $\rho(t,x,\lambda)$ satisfies the following hydrodynamic equations as 
\be\label{eq_GHD_1}
\partial_t\rho+\partial_x(v^\text{eff}\rho)+\partial_\lambda\left(\frac{\partial_t \alpha f^\text{dr}+\partial_x \alpha \Lambda^\text{dr}}{(\partial_\lambda p)^\text{dr}} \rho\right)=0\, 
\ee
where we dropped the space-time dependence to lighten the notation. In Eq. \eqref{eq_GHD_1} 
$v^{\text{eff}}$ is the dressed velocity of the quasiparticles. 
Only first-order derivatives appear, implying that the equation is invariant under 
the rescaling $(t,x)\to(A t,Ax)$, with $A\in\mathbb{R}^+$.
For a space-time homogeneous dynamics ($\partial_x\alpha=\partial_t\alpha=0$), the standard GHD equations are obtained \cite{transportbertini,hydrodoyon1}.
The forces $f$ and $\Lambda$ are obtained by solving 
\be\label{eq_force_1}
f(\lambda)=-\partial_\alpha p(\lambda)+\int \frac{\dd\mu}{2\pi}\partial_\alpha\Theta(\lambda-\mu)(\partial_\mu p)^\text{dr}\vartheta(\mu)\, ,
\ee
\be\label{eq_force_2}
\Lambda(\lambda)=-\partial_\alpha \epsilon(\lambda)+\int \frac{\dd\mu}{2\pi}\partial_\alpha \Theta(\lambda-\mu)(\partial_\mu \epsilon)^\text{dr}\vartheta(\mu) \, .
\ee
Here $\vartheta=2\pi
\rho/(\partial_\lambda p)^\text{dr}$ is the filling function.
As usual in GHD, Eq. \eqref{eq_GHD_1} has a clear semiclassical interpretation: $\rho(t,x,\lambda)$ locally describes the phase-space density of a 
collection of quasiparticles, moving with velocity $v^\text{eff}$ and subjected to force terms induced by the 
inhomogeneities, which can change the quasiparticles' rapidity. The force terms account for both 
single particle as well as collective effects.
The former are contained in the terms $\partial_\alpha p$ and $\partial_\alpha\epsilon$ in Eqs. (\ref{eq_force_1}-\ref{eq_force_2}). 
These are the energy-momentum changes of a single excitation of rapidity $\lambda$ induced by the inhomogeneities: 
the change in the dispersion relation causes the excitation to accelerate.
Force terms due to inhomogeneities have been previously derived in Ref. \cite{GHD3}, for spatially inhomogeneous potentials linearly coupled to the charge densities, which nevertheless cannot induce any inhomogeneity in the scattering data of the model.
In Ref. \cite{BasDeLu18}, slow magnetic flux changes in the XXZ spin chain have been studied. 
In both cases, only single-particle effects arise in the GHD equation and can now be regarded as a particular case of our more general findings.

\begin{figure}[t!]
\begin{center}
\includegraphics[width=0.4\columnwidth]{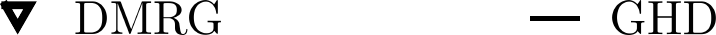}\\
\end{center}
\includegraphics[width=1\columnwidth]{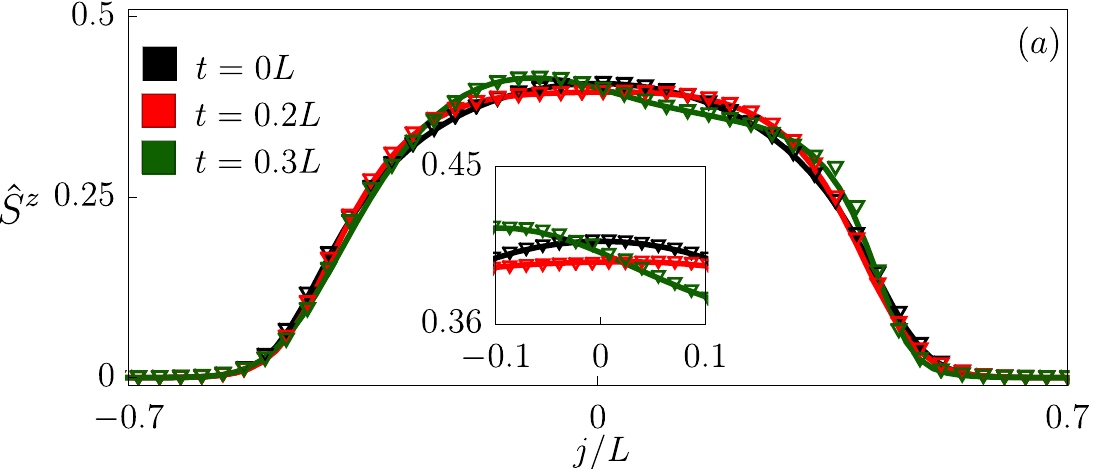}\\
\includegraphics[width=1\columnwidth]{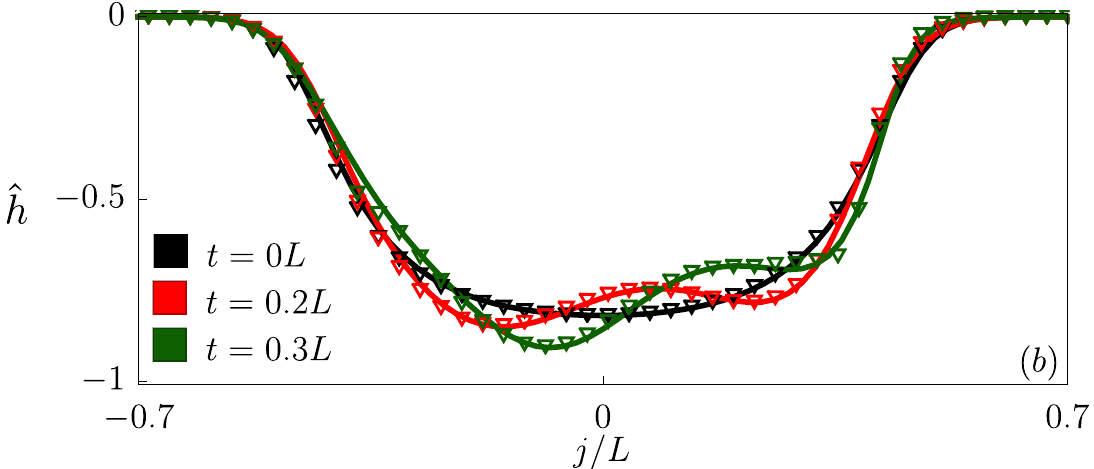}
\caption{\label{fig_num_check_XXZ}
Slow quench in the trapped XXZ spin chain. We apply the external static
magnetic field $B_j=-1-8(j/L)^2$, with $j$ the distance from the chain
center and $L$ its length. The initial state is a thermal one with $\beta=4$. 
We evolve the system with the XXZ chain with  $\Delta_j(t)=1.5+0.3\tanh(3t/L)\sin(4\pi(j-t)/L)$. 
$(a)$ Profile of the local magnetization $\hat S_j^z$ as a function of
$j/L$ and several times. The curves are GHD results. The symbols
are tDMRG simulations for a chain with $L=128$, and are 
in good agreement with the GHD. The inset shows
a zoom around the center of the system. $(b)$ Profile of the
local energy density $\hat h_j=\hat{S}^x_j\hat{S}^x_{j+1}+
\hat{S}^y_j\hat{S}^y_{j+1}+\Delta_j(t)\hat{S}_j^z\hat{S}^z_{j+1}-\Delta_j(t)/4$.
}
\end{figure}

The integrals in Eqs. \eqref{eq_force_1}-\eqref{eq_force_2} are entirely due to collective 
behaviors and have never been derived in previous studies.
Because of modifications in the interparticle interactions, encoded in the scattering 
phase $\Theta$, the excitations experience force fields caused by the surrounding 
particles.

For spatial-homogeneous interactions, i.e. $\partial_x\alpha=0$, we are able to 
derive Eq.~\eqref{eq_GHD_1} for rather generic integrable models \cite{suppl}. 
In the presence of spatial inhomogeneities, thus $\partial_x\alpha\ne0$, Eq. \eqref{eq_GHD_1} is derived in the presence of Lorentz invariance \cite{suppl} and in 
Galilean invariant models through a non relativistic limit \cite{KoMuTr10,KoMuTr09,BaMuDeLu16,BaMuDeLu17} (see SM \cite{suppl}). 
Outside of the mentioned cases, we present Eq.~\eqref{eq_force_2} as a conjecture, although well supported by numerical evidence (see Fig. \ref{fig_num_check_XXZ}).
As a further nontrivial check, thermal states are shown to be steady states of the GHD equation~\eqref{eq_GHD_1} with $\partial_x\alpha\ne 0$ \cite{suppl}.

We stress that, in order to have a weakly varying (locally) integrable model, a smooth dependence of $\hat{\bf h}(x,\alpha)$ \eqref{eq_inh_H} 
on the coupling does not suffice: the whole set of (quasi-)local charges must be smooth as a function of $\alpha$. 
For example, our method cannot be applied to interaction changes in the XXZ spin chain with 
$|\Delta|<1$, which has a fractal dependence on the coupling \cite{taka}.

\paragraph{Applications and numerical checks. ---}

We now show the wide applicability of our results. 
GHD equations are numerically solved according with the method described in the SM \cite{suppl}, where we also present a short summary of the TBA of the models here investigated.
In Fig. \ref{fig_LL} we show a possible application to an experimentally relevant setup, 
namely a (slow) interaction quench in the interacting Bose gas \cite{LiLi63,LiLi63_bis}. 
We mention that there are no alternative analytical and numerical methods to address this 
type of protocols. 
Closely related setups have already been experimentally addressed~\cite{FaCaJoBo16}. 

The Hamiltonian of the Lieb-Liniger model reads
$\hat{H}=\int \dd x\, \{\partial_x\hat{\psi}^\dagger\partial_x\hat{\psi}+c(t)
(\hat{\psi}^\dagger)^2(\hat{\psi})^2+V(x)\hat{\psi}^\dagger\hat{\psi}\}$, with $[\hat{\psi}(x),\psi^\dagger(y)]=\delta(x-y)$. 
The gas is loaded in a harmonic trap in a low-temperature state, the interaction $c(t)>0$ is then slowly 
increased. This induces a non trivial evolution of the quasiparticle densities, which are reported in 
Fig.~\ref{fig_LL} $(a.1)$.
As the interparticle repulsion is increased quasiparticles increase their rapidity $\lambda$ 
(reflected in the stretching of the initial blob along the vertical direction) and escape from the 
center of the trap. The local total density $n(t,x)=\int \dd\lambda\, \rho(\lambda)$  of the 
quasiparticles   is shown in Fig. \ref{fig_LL} $(a.2)$. 
Interestingly, the quench induces a breathing mode, 
which is long-lived in harmonic potentials \cite{FaCaJoBo16}. This is clear from Fig. \ref{fig_LL} $(a.3)$, 
where we show the density $n$ in the center of the trap as a function of time. 
In Fig. \ref{fig_LL} $(b.1)\, (b.3)$ we focus on the slow quench in an anharmonic trap. 
As it is clear from Fig.~\ref{fig_LL} $(b.1)$ the anharmonicity causes a 
spiral motion in the filling which develops a fractal structure as times passes~\cite{CDDK17,CaoBuMo18}, which is smoothed due to the discretization used to solve the GHD equation~\eqref{eq_GHD_1}.
Now a much faster relaxation is observed as compared with the harmonic case, due to dephasing \cite{CDDK17,BaCoSo17}.

\begin{figure}[t!]
\includegraphics[width=0.5\columnwidth]{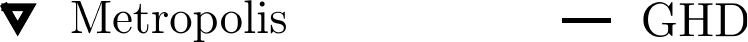}\ \\ \, \ \\
\includegraphics[width=1\columnwidth]{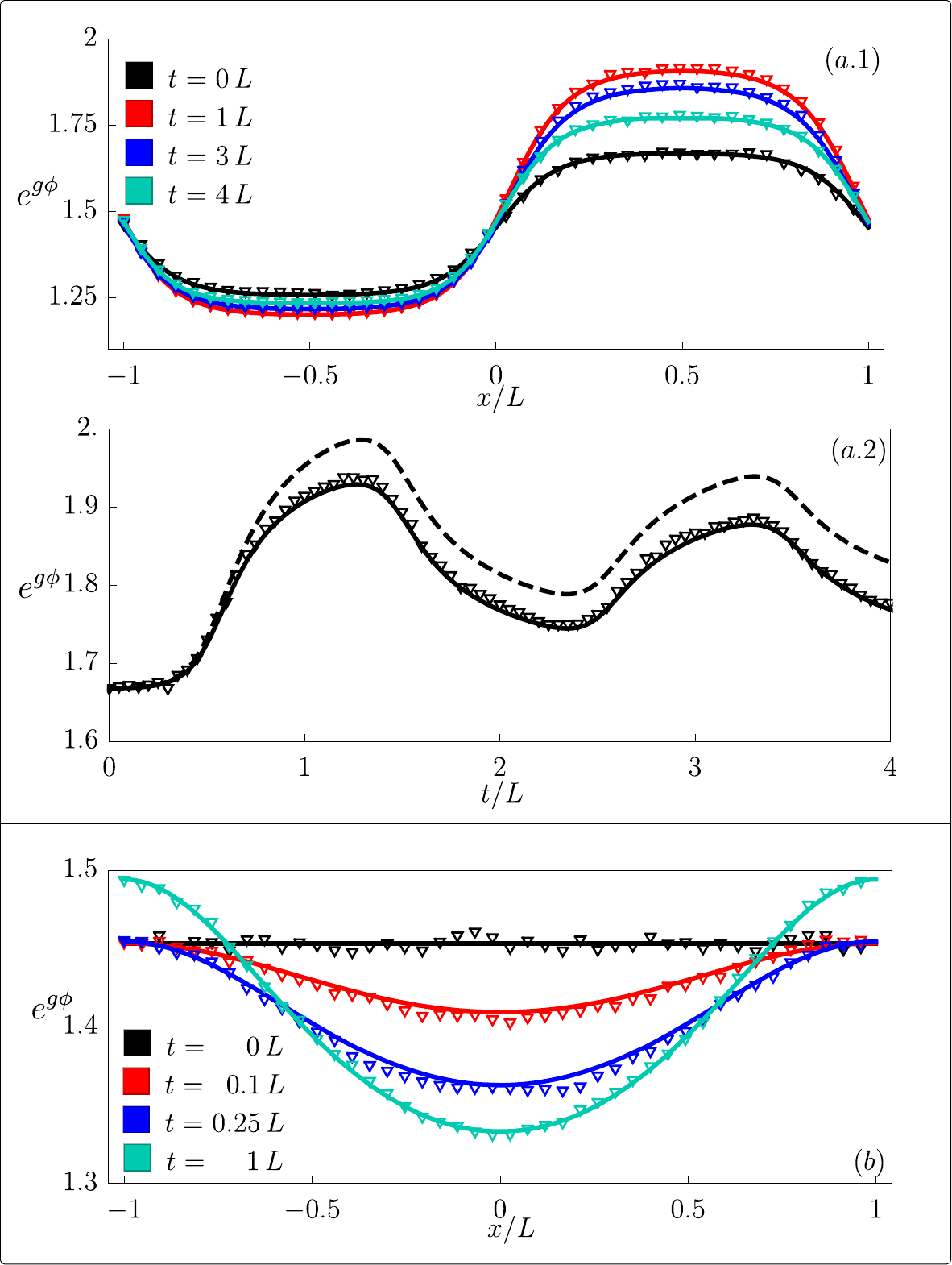}
\caption{\label{fig_num_check_ShG}
Classical sinh-Gordon model: GHD results compared
 with classical Monte Carlo simulations. Panels $(a.1)-(a.2)$: the system is prepared in an 
 inhomogeneous thermal state with inverse temperature $\beta(x)=1.25+0.25 
 \tanh[2\sin(2\pi x/L)]$, with $L$ the system length. 
 The system is evolved with the sinh-Gordon Hamiltonian with inhomogeneous coupling 
 $g(x)=1.5+0.5 \tanh[2\sin(2\pi x/L)]$ and $m=1$.
 $(a.1)$. Profile of the vertex operator $\langle e^{g\phi}\rangle$ as a function of
$x/L$ at different times. The curves are GHD predictions. Symbols are
classical Monte Carlo simulations for $L=30$.
 $(a.2)$. Vertex operator $\langle e^{g\phi}\rangle$ at $x=L/2$
 as a function of time. Now the continuous line is the GHD prediction.
 The dashed line is the result obtained by ignoring the
 the force fields, i.e. $\Lambda=0$ in Eq.~\eqref{eq_GHD_1},
 which is inaccurate, as expected.
Panel $(b)$: the system is prepared in an homogeneous thermal ensemble with $m=1$, $g=1.5$ and $\beta=1.25$. At $t>0$ we vary the mass $m\to m(t,x)=m+\Delta m\,  4 t /L(\cos(2\pi x/L)-1)$ for $t\le L/4$ and $m(t,x)=m+\Delta m (\cos(2\pi x/L)-1)$ for $t>L/4$, we choose $\Delta m=0.25$ and the interaction $g$ is kept constant. The profiles of the vertex operator at different times are displayed.
}
\end{figure}

In Fig.~\ref{fig_num_check_XXZ} we focus on the XXZ spin chain with Hamiltonian 
$\hat{H}=\sum_{j=-L}^L \{\hat{S}^x_j\hat{S}^x_{j+1}+\hat{S}^y_j\hat{S}^y_{j+1}+\Delta_j(t)\hat{S}_j^z\hat{S}^z_{j+1}+B_j\hat{S}^z_j\}$, where
$\hat{S}_j^\alpha$ are standard spin$-1/2$ operators.
The system is initialized in a confining magnetic field and in a low temperature thermal state, with a uniform interaction $\Delta_j>1$. Then, $\Delta_j$ is slowly changed with time in the form of a traveling wave (see Fig.~\ref{fig_num_check_XXZ}). 
In Fig.~\ref{fig_num_check_XXZ} we compare the GHD predictions for the local magnetization and the local energy density with tDMRG simulations \cite{itensor,greg,metts}, finding excellent agreement. tDMRG methods suffer strong limitations in the accessible time scale, therefore we revert to the classical world to explore longer time scales.
In Fig. \ref{fig_num_check_ShG} we benchmark the GHD in the classical sinh-Gordon 
model (see Ref. \cite{DeLucaMuss16} and \cite{BDWY17} for the TBA). The model describes a scalar 
field $\phi$ with Lagrangian $\mathcal{L}=\int \dd x\,\{\partial_\mu\phi\partial^\mu \phi/2-(m/g)^2[\cosh(g\phi)-1]\}$, 
with $m$ the mass and $g$ the interaction parameter. 
We consider two protocols. In the first case (panels $(a.1)-(a.2)$) the system is initially prepared in a 
thermal state with an inhomogeneous temperature profile. 
The system is then evolved with the sinh-Gordon Hamiltonian with 
inhomogeneous coupling $g\to g(x)$.
Such an inhomogeneity does not affect the single-particle dispersion law (see SM \cite{suppl}), providing an ideal benchmark to test the collective effects in 
Eq.(\ref{eq_force_1}-\ref{eq_force_2}).
Symbols are Monte Carlo data \cite{BDWY17},  whereas the lines are the GHD results.
In $(a.1)$ we show the expectation 
value of the vertex operator $e^{g\phi}$ as a function of position $x/L$ for different times,
in $(a.2)$ we plot $e^{g\phi}$ at $x=L/2$ as a function of time. The agreement with 
the GHD is spectacular.
The dashed line is the GHD result neglecting the 
collective effects, i.e. the integrals in the right hand side in Eq.~\eqref{eq_force_1}, which clearly have a crucial role.
In the second protocol (panel $(b)$), we start from a homogeneous thermal ensemble, then a mass inhomogeneity is slowly activated with a linear ramp in time up to $t=L/4$ and then kept constant.  Mass changes do not affect the scattering data \cite{suppl}, implying that only single-particle effects in Eq.(\ref{eq_force_1}-\ref{eq_force_2}) play a role. Despite the form of the force terms, mass inhomogeneities do not belong to the class of inhomogeneities described in Ref. \cite{GHD3} and thus provide a nontrivial benchmark of our findings.

\paragraph{Conclusions and outlook.---}

The success of hydrodynamic approaches is hard to overestimate. 
GHD merges the hydrodynamic framework with integrability, providing unprecedented levels 
of accuracy in describing out-of-equilibrium systems. 
In this Letter we extended the reach of this program, providing hydrodynamic equations which 
account for arbitrary (smooth) inhomogeneities in the couplings and state.
Several interesting questions are left out for the future. Our analysis holds true when the model 
has a smooth dependence on the inhomogeneous coupling, but there could be special points (or regions) 
where this hypothesis breaks down. Understanding the behavior of protocols overcoming such 
special points is surely a compelling quest, which can unveil a rich phenomenology 
(see Ref. \cite{BasDeLu18} for a closely related problem). Including higher order 
corrections in the derivative expansion at the root of Eq. \eqref{eq_GHD_1} is another important direction. 
Finally, it is important to devise numerical schemes based on molecular dynamics, such as the flea gas~\cite{GHD6}, 
to simulate the GHD equations~\eqref{eq_GHD_1}.

\begin{acknowledgements}
\paragraph{Acknowledgments. --- } A.B. is indebted with A. De Luca for useful discussions and work on related subjects.
The authors acknowledge the support from the European
Research Council under ERC Advanced Grant
 No. 743032 DYNAMINT.

\end{acknowledgements}


\onecolumngrid
\newpage

\setcounter{equation}{0}            
\setcounter{section}{0}             
\renewcommand\thesection{\Alph{section}}    
\renewcommand\thesubsection{\arabic{subsection}}    
\renewcommand{\thetable}{S\arabic{table}}
\renewcommand{\theequation}{S\arabic{equation}}
\renewcommand{\thefigure}{S\arabic{figure}}
\setcounter{secnumdepth}{1}  

\begin{center}
{\Large Supplemental Material\\
\titleinfo
}
\end{center}

The Supplementary Material provides some technical analysis which complement the main text. It is organized as it follows

\begin{enumerate}
\item Section \ref{SM_sec_GHDeq} goes through a detailed derivation of the GHD equation presented in the main text.
\item Section \ref{SM_sec_num} deals with the numerical solution of the GHD equation, presenting an algorithm with $O(\dd t^2)$  precision.
\item Section \ref{SM_sec_TBA} briefly presents the details of the TBA of the models we analyzed, additional details on the numerical simulations are given as well.
Section \ref{suppl_NR_sec} derives the GHD of the Interacting Bose gas, viewing the latter as the non relativistic limit of the sinh-Gordon model.
\end{enumerate}

\section{Derivation of the GHD equations}
\label{SM_sec_GHDeq}

Following the original references Ref. \cite{S_transportbertini,S_hydrodoyon1}, we assume local relaxation to a (weakly) inhomogeneous GGE, yet to be determined. The (local) GGE is unambiguously fixed by the expectation value of all the (quasi-) local charges.
Let us consider the family of integrable models described by the parameter-dependent Hamiltonian $\hat{H}(\alpha)$ out of which we constructed the inhomogeneous Hamiltonian. Similarly, we consider the parameter-dependent (quasi-)local charges $\hat{\mathcal{Q}}_j(\alpha)=\int \dd x\, \hat{\bf q}_j(x,\alpha)$ of the homogeneous system, then construct
\be
\hat{\mathcal{Q}}_j=\int \dd x\, \hat{\bf q}_j(x,\alpha(t,x))\, . \label{S_inh_ch}
\ee

The operator \eqref{S_inh_ch} pointwise resembles a local charge, but it is not conserved anymore due to the inhomogeneity, which breaks integrability on a large scale. Nevertheless, the knowledge of $\langle \hat{\bf q}_j(x,\alpha(t,x))\rangle$ for any $j$ fixes the local GGE.
Let us write the Heisenberg equation of motion for the local density after an infinitesimal evolution $t\to t+\dd t$. Let $\hat{\bf q}^\text{H}_j(t,x,\alpha(t,x))$ be the local charge density in the Heisenberg picture, its time variation receives a contribution from the Hamiltonian evolution and one from the parametric change
\be
\hat{\bf q}^\text{H}_j(t+\dd t,x,\alpha(t+\dd t,x))=e^{i\dd t \hat{H}(t)}\hat{\bf q}_j(x,\alpha(t+\dd t,x))e^{-i\dd t \hat{H}(t)}\, .
\ee

Above, we added a label "$t$" to the Hamiltonian to stress its explicit time dependence. Expanding at $O(\dd t)$ we find
\be
\partial_t \hat{\bf q}^\text{H}_j(t,x,\alpha(t,x))=\partial_t\alpha(t,x)\partial_\alpha \hat{\bf q}_j(x,\alpha(t,x)+i\int \dd y\, [\hat{\bf h}(y,\alpha(t,y)),\hat{\bf q}_j(x,\alpha(t,x))])\,.
\ee

Above, $\hat{\text{\bf h}}$ is the local Hamiltonian density. We further manipulate the integral expanding $\alpha(t,y)\simeq \alpha(t,x)+(y-x)\partial_x \alpha(t,x)+...$ . Higher derivatives can be neglected in the limit of smooth variations.
\begin{multline}
i\int \dd y\, [\hat{\bf h}(y,\alpha(t,y)),\hat{\bf q}_j(x,\alpha(t,x))]=i\int \dd y\, [\hat{\bf h}(y,\alpha(t,x)),\hat{\bf q}_j(x,\alpha(t,x))]+\\ i\int \dd y\, (x-y) \partial_x\alpha(t,x)[\partial_\alpha\hat{\bf h}(y,\alpha(t,x)),\hat{\bf q}_j(x,\alpha(t,x))]+...\, .
\end{multline}
In the first term of the r.h.s., the parameter $\alpha$ is constant and we can compute the expression within the homogeneous case, resulting in the divergence of the proper current operator of the homogeneous model.
Therefore, we get
\be
\partial_t \hat{\bf q}_j(x,\alpha(t,x))+\partial_x \hat{\bf j}_j(x,\alpha(t,x))-\partial_t\alpha(t,x)\partial_\alpha \hat{\bf q}_j(x,\alpha(t,x)-\partial_x\alpha\hat{\Phi}(t,x)=0\,.
\ee
Above, we drop further orders in the derivative expansion of $\alpha$ (negligible at first order in the infinitely smooth limit) and defined
\be
\hat{\Phi}_j(t,x)=i\int \dd y\, (x-y) [\partial_\alpha\hat{\bf h}(y,\alpha(t,x)),\hat{\bf q}_j(x,\alpha(t,x))]\, .
\ee

From the Heisenberg equation of motion, we want now to move to expectation values and invoke local relaxation to the inhomogeneous GGE. To this aim, we approximate the expectation values of space-time derivatives of charges and currents with the derivatives of the expectation values on the inhomogeneous GGE \cite{S_transportbertini,S_hydrodoyon1}, i.e.
\be
\langle \partial_t \hat{\bf q}^\text{H}_j(t,x,\alpha(t,x))\rangle\simeq \partial_t  \langle\hat{\bf q}_j(x,\alpha(t,x))\rangle_{\text{GGE}(t,x)}\,,\hspace{2pc}\langle \partial_x \hat{\bf j}^\text{H}_j(t,x,\alpha(t,x))\rangle\simeq \partial_x \langle\hat{\bf j}_j(x,\alpha(t,x))\rangle_{\text{GGE}(t,x)}\,.
\ee

Enforcing this approximation, we are finally lead to the (infinite set of) equations
\be\label{S_eq_expeq}
\partial_t \langle\hat{\bf q}_j\rangle+\partial_x \langle\hat{\bf j}_j\rangle-\partial_t\alpha\langle\partial_\alpha \hat{\bf q}_j\rangle-\partial_x \alpha\langle\hat{\Phi}_j\rangle=0\,.
\ee
For simplicity, we drop the explicit dependence of the various operators and the expectation values are meant to be taken over the inhomogeneous GGE at the point of interest. Enforcing these equations on the complete set of charges, we aim for an equation for the root density: in this perspective, we need the expectation value of the various operators. The charge expectation value is the simplest: again, we focus on a single type of excitation, but everything we say is readily generalized to multiparticle species.
\be\label{S_eq_ch}
\langle\hat{\bf q}_j(x,\alpha(t,x))\rangle_{\text{GGE}(t,x)}=\int \dd \lambda\,  q_j(\lambda,\alpha(t,x)) \rho(t,x,\lambda)\, .
\ee
Above, $q_j$ is the charge eigenvalue in which we made explicit the dependence on the inhomogeneous coupling. Taking the time derivative we have
\be\label{S_eq_ch_dt}
\partial_t\langle\hat{\bf q}_j\rangle=\int \dd \lambda\,  \partial_t\alpha\partial_\alpha q_j(\lambda) \rho(\lambda)+ q_j(\lambda) \partial_t\rho(\lambda)\, .
\ee
Above, we suppress the explicit space-time dependence for the seek of a lighter notation.
The expectation value of the current is less trivial and it has been only recently computed \cite{S_transportbertini,S_hydrodoyon1}, making possible the first formulation of GHD (which lacks the terms $\propto \partial\alpha$ in Eq. \eqref{S_eq_expeq})
\be\label{S_eq_j}
\langle\hat{\bf j}_j(x,\alpha(t,x))\rangle_{\text{GGE}(t,x)}=\int \dd \lambda\,  q_j(\lambda,\alpha(t,x)) v^\text{eff}(t,x,\lambda)\rho(t,x,\lambda)\, .
\ee
Above, the effective velocity has a space-time dependence due both to the dressing and to the parametric dependence on the coupling $\alpha$.
We write its spatial derivative as it follows (below, the $(t,x)$ dependence is neglected in the notation since no ambiguities arise)
\be\label{S_eq_j_dx}
\partial_x\langle\hat{\bf j}_j(x,\alpha(t,x))\rangle_{\text{GGE}(t,x)}=\int \dd \lambda\,  \partial_x\alpha\partial_\alpha q_j(\lambda) v^\text{eff}(\lambda)\rho(\lambda)+ q_j(\lambda) \partial_x\left(v^\text{eff}(\lambda)\rho(\lambda)\right)\, .
\ee

Computing the remaining terms is an open problem, which we managed to partially solve. Indeed, $\langle\partial_\alpha \hat{\bf q}_j\rangle$ can be exactly computed through a generalization of the Hellmann-Feynman theorem: we postpone the derivation to the end of this Section and, for the time being, just quote the result. On arbitrary GGEs we have
\be\label{S_eq_HF}
\langle \partial_\alpha \hat{\bf q}_j(x,\alpha)\rangle=\int \dd\lambda\, \partial_\alpha q_j(\lambda,\alpha)\rho_j(\lambda)+ \frac{1}{2\pi}f^\text{dr}(\lambda) \partial_\lambda q_j(\lambda,\alpha)\vartheta(\lambda)\,,
\ee
where $\vartheta$ is the filling and the function $f$ is defined in the main text Eq. (5) ($f$ has a parametric dependence on $\alpha$ which we drop for the seek of simplicity).
It is useful to perform an integration by parts: assuming that the boundary terms in the integral vanish (it is usually the case, see however Ref. \cite{S_BasDeLu18}) we have
\be\label{S_eq_HF_parts}
\langle \partial_\alpha \hat{\bf q}_j\rangle=\int \dd\lambda\, \partial_\alpha q_j(\lambda)\rho_j(\lambda)- \frac{1}{2\pi}q_j(\lambda)\partial_\lambda\left(f^\text{dr}(\lambda) \vartheta(\lambda)\right)\, .
\ee

The knowledge of $\langle \partial_\alpha \hat{\bf q}_j\rangle$ is enough to deal with those protocols where the dynamics is time-dependent, but homogeneous, i.e. $\partial_x \alpha=0$. Notice that we do not need to require the homogeneity of the state.
However, we want to provide an answer for arbitrary inhomogeneities, thus $\partial_x\alpha\ne 0$.

Computing $\langle \hat{\Phi}_j\rangle$ is much more complicated and we did not succeed in providing a first-principle derivation. However, invoking some reasonable assumptions which we discuss later on, the natural ansatz for the GHD equation presented in the main text emerges. Lately, the ansatz can be proven in presence of Lorentz invariance.

Let us plug (\ref{S_eq_ch_dt}-\ref{S_eq_j_dx}-\ref{S_eq_HF_parts}) into \eqref{S_eq_expeq} leaving $\langle \hat{\Phi}_j\rangle$ implicit
\be\label{S_eq_int_GHD}
\int \dd \lambda\,   q_j(\lambda) \left[\partial_t\rho(\lambda)+\partial_x\left(v^\text{eff}(\lambda)\rho(\lambda)\right)+ \frac{\partial_t\alpha}{2\pi}\partial_\lambda\left(f^\text{dr}(\lambda) \vartheta(\lambda)\right)\right]+\partial_x\alpha\left[-\langle \hat{\Phi}_j\rangle+\int \dd \lambda\,  \partial_\alpha q_j(\lambda) v^\text{eff}(\lambda)\rho(\lambda)\right]=0\, .
\ee

It is convenient to stop for a moment and consider $\partial_x\alpha=0$. In this case, following \cite{S_transportbertini,S_hydrodoyon1}, we invoke the completeness of the charges and replace the infinite set of integral equations (holding true for any charge $\hat{\bf q}_j$) with a differential equation for $\rho$, obtained posing to $0$ the term in Eq. \eqref{S_eq_int_GHD} proportional to $q_j(\lambda)$.
The presence of the unknown term $\langle \hat{\Phi}_j\rangle$ prevents us from straightforwardly apply the same reasoning to the case $\partial_x\alpha\ne 0$. However, we assume the existence of a GHD equation for the root density which, compared with the case $\partial_x\alpha=0$, adds a yet unknown contribution
\be\label{S_eq_rho_hydro_chi}
\partial_t\rho+\partial_x\left(v^\text{eff}\rho\right)+ \partial_t\alpha\frac{1}{2\pi}\partial_\lambda\left(f^\text{dr}\vartheta\right)+\partial_x \alpha\,  \chi=0\, .
\ee
Above, $\chi(t,x,\lambda)$ is due to the second term in \eqref{S_eq_int_GHD}. Invoking the locality of the GHD equation, $\chi(t,x,\lambda)$ must be completely determined by the model at $(t,x)$, i.e. by $\rho(t,x)$ and $\alpha(t,x)$. It cannot contain space or time derivatives neither of the root density or of the coupling, since these terms would be next-to-leading order in the weakly-inhomogeneity approximation.
The problem is now reduced to the determination of $\chi$. To this aim, we convert Eq. \eqref{S_eq_rho_hydro_chi} into an equation for the filling function $\vartheta$, namely
\be\label{S_eq_filldef}
\vartheta(\lambda)=2\pi \frac{\rho(\lambda)}{(\partial_\lambda p)^\text{dr}}\, .
\ee

The experience gained from the previous literature (see e.g. Ref. \cite{S_transportbertini,S_hydrodoyon1,S_GHD3,S_BasDeLu18}) teaches us that simple equations for the filling should be expected: this rewriting leads us to a very natural ansatz for $\chi$.
In order to reach the desired equation, we start computing the time derivative of the dressed momentum derivative $\partial_t (\partial_\lambda p)^\text{dr}$. From the definition of the dressing we have (again, we suppress the explicit $(t,x)$ dependence if no ambiguities arise)
\be
\partial_t\left[\frac{(\partial_\lambda p)^\text{dr}}{2\pi}\right]=\frac{\partial_t\alpha}{2\pi}\partial_\lambda\left( p(\lambda)-\int\dd \mu\, \partial_\alpha\Theta(\lambda-\mu) \rho(\mu)\right)-\int \frac{\dd \mu}{2\pi}\partial_\lambda\Theta(\lambda-\mu) \partial_t\rho(\mu)\, .
\ee
The first term in round brackets is readily identified with $-f(\lambda)$, as defined in the main text Eq. (5). In the term where $\partial_t\rho$ appears, we take advantage of the hydrodynamic equation \eqref{S_eq_rho_hydro_chi}. Furthermore, we use the following identities
\begin{multline}
\int \frac{\dd \mu}{2\pi}\partial_\lambda\Theta(\lambda-\mu) \partial_x(v^\text{eff}(\mu)\rho(\mu))=\partial_x\left[\int \frac{\dd \mu}{2\pi}\partial_\lambda\Theta(\lambda-\mu) v^\text{eff}(\mu)\rho(\mu)\right]-\partial_x\alpha\partial_\lambda\left[\int \frac{\dd \mu}{2\pi}\partial_\alpha\Theta(\lambda-\mu) v^\text{eff}(\mu)\rho(\mu)\right]=\\
=-\frac{1}{2\pi}\partial_x\left[(\partial_\lambda \epsilon)^\text{dr}-(\partial_\lambda \epsilon)\right]-\partial_x\alpha\partial_\lambda\left[\int \frac{\dd \mu}{2\pi}\partial_\alpha\Theta(\lambda-\mu) v^\text{eff}(\mu)\rho(\mu)\right]
\end{multline}
and
\begin{multline}
\int \frac{\dd \mu}{2\pi}\partial_\lambda\Theta(\lambda-\mu) \partial_\mu\left(f^\text{dr}(\mu)\vartheta(\mu)\right)=-\int \frac{\dd \mu}{2\pi}\partial_\lambda\partial_\mu\Theta(\lambda-\mu) f^\text{dr}(\mu)\vartheta(\mu)=\\
=\partial_\lambda\left[\int \frac{\dd \mu}{2\pi}\partial_\lambda\Theta(\lambda-\mu) f^\text{dr}(\mu)\vartheta(\mu)\right]=-\partial_\lambda \left[f^\text{dr}(\lambda)-f(\lambda)\right]\, .
\end{multline}
Above, we integrated by parts assuming zero contribution from the boundary terms, use the symmetry of the kernel and finally the definition of the dressing.
Collecting the various terms we can write
\begin{multline}
\partial_t\left[\frac{(\partial_\lambda p)^\text{dr}}{2\pi}\right]=-\frac{\partial_t\alpha}{2\pi}\partial_\lambda f^\text{dr}(\lambda)-\frac{1}{2\pi}\partial_x(\partial_\lambda \epsilon)^\text{dr}\\
-\frac{1}{2\pi}\partial_x\alpha\left\{-\partial_\lambda\left[\partial_\alpha\epsilon(\lambda)-\int \frac{\dd \mu}{2\pi}\partial_\alpha\Theta(\lambda-\mu)\vartheta(\mu)(\partial_\mu \epsilon)^\text{dr} \right]-\int \dd \mu\,\partial_\lambda\Theta(\lambda-\mu) \chi(\mu)\right\}\, .
\end{multline}

Using now this last result and the definition of the filling Eq. \eqref{S_eq_filldef}, we finally reach the following hydrodynamic equation

\begin{multline}\label{S_eq_theta_hydro_chi}
\partial_t\vartheta(\lambda)+v^\text{eff}(\lambda)\partial_x\vartheta(\lambda)+\frac{\partial_t\alpha f^\text{dr}(\lambda)}{(\partial_\lambda p)^\text{dr}}\partial_\lambda\vartheta(\lambda)+\\
+\frac{\partial_x\alpha}{(\partial_\lambda p)^\text{dr}}\Bigg\{
\vartheta(\lambda)\partial_\lambda\left[\partial_\alpha\epsilon(\lambda)-\int \frac{\dd \mu}{2\pi}\partial_\alpha\Theta(\lambda-\mu)\vartheta(\mu)(\partial_\mu \epsilon)^\text{dr} \right]+2\pi\chi(\lambda)+\vartheta(\lambda) \int \dd \mu\,\partial_\lambda\Theta(\lambda-\mu) \chi(\mu)\Bigg\}=0
\end{multline}

This is how further we can go without any additional hypothesis on $\chi$ or symmetries of the system.
Notice that the contribution proportional to $\partial_t\alpha$, passing from Eq. \eqref{S_eq_rho_hydro_chi} to Eq. \eqref{S_eq_theta_hydro_chi}, retains a very simple form, while the term $\partial_x\alpha$ looks strangely complicated.
Inspired by the $\propto \partial_t \alpha$ term, we make the following ansatz
\be
\chi(\lambda)=\frac{1}{2\pi}\partial_\lambda (\Lambda^\text{dr}(\lambda) \vartheta(\lambda))\hspace{2pc} \textbf{ansatz}\, ,
\ee
with $\Lambda$ a still unknown function, Eq. \eqref{S_eq_theta_hydro_chi} is then greatly simplified
\begin{multline}
\partial_t\vartheta(\lambda)+v^\text{eff}(\lambda)\partial_x\vartheta(\lambda)+\frac{\partial_t\alpha f^\text{dr}(\lambda)+\partial_x\alpha\Lambda^\text{dr}(\lambda)}{(\partial_\lambda p)^\text{dr}}\partial_\lambda\vartheta(\lambda)+\\
+\frac{\partial_x\alpha \vartheta(\lambda)}{(\partial_\lambda p)^\text{dr}}
\partial_\lambda\left[\Lambda(\lambda)+\partial_\alpha\epsilon(\lambda)-\int \frac{\dd \mu}{2\pi}\partial_\alpha\Theta(\lambda-\mu)\vartheta(\mu)(\partial_\mu \epsilon)^\text{dr} \right]=0\, .
\end{multline}
At this point, it is very tempting to assume the identification
\be\label{S_eq_Lambda_ansatz}
\Lambda(\lambda)=-\partial_\alpha\epsilon(\lambda)+\int \frac{\dd \mu}{2\pi}\partial_\alpha\Theta(\lambda-\mu)\vartheta(\mu)(\partial_\mu \epsilon)^\text{dr} \, , \hspace{2pc} \textbf{ansatz}
\ee
i.e. Eq. (6) of the main text. This immediately enforces the hydrodynamic equation for the filling
\be\label{S_eq_GHD_2}
\partial_t\vartheta+v^\text{eff}\partial_x\vartheta+\frac{\partial_t \alpha f^\text{dr}+\partial_x \alpha \Lambda^\text{dr}}{(\partial_\lambda p)^\text{dr}} \partial_\lambda\vartheta=0\, ,
\ee

which is equivalent to Eq. (4) of the main text.
Apart from the appealing formal structure, nontrivial checks can be performed.
In the main text we provided numerical benchmarks of our result in a variety of contexts, finding excellent agreement. Furthermore, we can explicitly check that thermal states in the local density approximation are steady states of the hydrodynamic equation, as it should be. This check is performed in the next short subsection.

Lastly, we provide a derivation of our ansatz in Lorentz invariant models, from which we can assess galilean invariant models through proper non relativistic limits.

\subsection{Check: thermal states are steady states of the GHD equation}
As long as we are interested in GGEs described by thermal states, their filling is best parametrized in terms of the effective energy $\varepsilon$ \cite{S_taka} as it follows
\be
 \vartheta(\lambda)=\frac{1}{e^{\varepsilon(\lambda)}+1}\,.
\label{eq:filling}
\ee
The effective energy satisfies the following integral equation
\be
\varepsilon(\lambda)=\beta \epsilon(\lambda)+\int_{-\infty}^{\infty}\frac{{\rm d}\mu}{2\pi}\partial_\lambda\Theta(\lambda-\mu)\log\left(1+e^{-\varepsilon(\mu)}\right)\,.
\label{eq:thermal_TBA}
\ee
Or, equivalently
\be\label{eq:thermal_TBA_2}
\varepsilon(\lambda)=\beta \epsilon(\lambda)-\int_{-\infty}^{\infty}\frac{{\rm d}\mu}{2\pi}\Theta(\lambda-\mu)\vartheta(\mu)\partial_\mu\varepsilon(\mu)\, .
\ee

We now consider an inhomogeneous system which is in a thermal state with inverse temperature $\beta$: within the local density approximation, the local GGE is fixed as per above where the energy eigenvalue has a parametric dependence on the position. Of course, such a state must be a steady state for the GHD equation.
Thus, we plug Eq. \eqref{eq:filling} in Eq. \eqref{S_eq_GHD_2} assuming $\partial_t \alpha=0$ (but keeping $\partial_x \alpha\ne 0$) and imposing $\partial_t\vartheta=0$.
We then reach the following equation for the effective energy
\be\label{eq_hydro_static}
(\partial_\lambda \epsilon)\partial_x\varepsilon+\partial_x \alpha \Lambda^\text{dr} \partial_\lambda\varepsilon=0\, .
\ee
Deriving the defining equation of the effective energy Eq. \eqref{eq:thermal_TBA_2} in the rapidities we readily get $\partial_\lambda \varepsilon=(\partial_\lambda \epsilon)^\text{dr}$. Instead, deriving with respect to the position we find $\partial_x \varepsilon =-\partial_x\alpha \Lambda^\text{dr}$. Thus, Eq. \eqref{eq_hydro_static} is satisfied.

\subsection{Derivation of the ansatz in the relativistic invariant case}

In addition to integrability we now assume the system to be relativistic invariant (we set the speed of light equal to unity, for simplicity). We start from the hydrodynamic equation in terms of the filling \eqref{S_eq_theta_hydro_chi}, but no hypothesis on the $\chi$ functions are made. We rewrite \eqref{S_eq_theta_hydro_chi} in a more compact way, collecting into an unknown function $w(\lambda)$ the $\propto \partial_x \alpha$ term
\be
(\partial_\lambda p)^\text{dr}\partial_t\vartheta(\lambda)+(\partial_\lambda \epsilon)^\text{dr}\partial_x\vartheta(\lambda)+\partial_t\alpha f^\text{dr}(\lambda)\partial_\lambda\vartheta(\lambda)+\partial_x\alpha \,w(\lambda)=0\, .
\ee
We now enforce relativistic invariance on the dispersion law, having $\epsilon(\lambda)=m \cosh\lambda$, $p(\lambda)=m\sinh\lambda $, with $m$ the mass of the fundamental excitation.
Therefore, it holds true
\be\label{S_eq_dis_lorentz}
\partial_\lambda\epsilon(\lambda)=p(\lambda)\,,\hspace{3pc}\partial_\lambda p(\lambda)=\epsilon(\lambda)\, .
\ee
We now construct the contravariant momentum $P^\mu(\lambda)=(\epsilon(\lambda),p(\lambda))$, furthermore we collect in an unique two component vector the force terms $F^\mu=(f^\text{dr}(\lambda)\partial_\lambda\vartheta(\lambda),w(\lambda))$. The hydrodynamic equation can be then rewritten as (sum over repeated indexes)
\be
(P^\mu)^\text{dr}\partial_\mu \vartheta+\partial_\mu \alpha F^\mu=0
\ee
Since $\vartheta$ is a scalar under Lorentz boosts, $(P^\mu)^\text{dr}$ inherits the same transformation properties of $P^\mu$. Therefore, $(P^\mu)^\text{dr}\partial_\mu \vartheta$ is a Lorentz scalar.
In order to complete the hydrodynamic equation to a Lorentz scalar, we are forced to require $F^\mu$ to be contravariant.

We can now use a Lorentz boost to fix $F^1$, using the knowledge of $F^0$ and the transformation properties under Lorentz boosts.
Let us consider a boost of velocity $v$, then $F^\mu\to (F^\mu)'$, in particular the first component
\be
(F^0)'= \gamma F^0-v \gamma F^1
\ee
with $\gamma=1/\sqrt{1+v^2}$. Using the definition of $f(\lambda)$, the identities \eqref{S_eq_dis_lorentz} and the tranformation properties of energy and momentum, from the above equation we can read $F^1$, which turns out to be
\be
F^1(\lambda)=\Lambda^\text{dr}(\lambda)\partial_\lambda \vartheta(\lambda)\, ,
\ee
with
\be
\Lambda(\lambda)=-\partial_\alpha\epsilon(\lambda)+\int \frac{\dd \mu}{2\pi}\partial_\alpha\Theta(\lambda-\mu)\vartheta(\mu)p^\text{dr}(\mu) \, ,
\ee
i.e. Eq. \eqref{S_eq_Lambda_ansatz} specialized to the Lorentz-invariant case.

\subsection{Expectation value of the derivative of charges}
\label{S_subsec_HF}

During the derivation of the GHD equations, we postponed the proof of Eq. \eqref{S_eq_HF}, i.e. $\langle \partial_\alpha \hat{\bf q}_j(x,\alpha)\rangle$ computed on an arbitrary GGE. We now provide the proof through a suitable generalization of the Hellmann-Feynman theorem.
Firstly, we should take a step back from the thermodynamic limit and consider the system at finite size $L$, periodic boundary conditions (PBC) are assumed.
Let us consider a state $|\{\lambda_i\}_{i=1}^N\rangle$: due to the PBC, the rapidities must satisfy the Bethe-Gaudin equations \cite{S_taka}
\be\label{S_eq_BetheGaudin}
\frac{I_i}{L}=\frac{p(\lambda_i)}{2\pi}-\frac{1}{2\pi L}\sum_{j\ne i} \Theta(\lambda_i-\lambda_j)\, .
\ee
Above, $I_i$ are suitable integers.
Of course, we are lastly interested in the thermodynamic limit $N,L\to \infty$ ($N/L$ constant). In view of the representative state approach \cite{S_CaEs13,S_Ca16}, in the thermodynamic limit we can equivalently compute $\langle \partial_\alpha \hat{\bf q}_j(x,\alpha)\rangle$ on a single state rather than the whole GGE ensemble, provided that the root density associated with the representative state equals the GGE root density.

Rather than labeling the state with the rapidities, we use the Bethe integers. Moreover, we take advantage of the homogeneity of the GGE and compute the derivative of the whole charge, rather than its density
\be
\langle \{I_i\}_{i=1}^N|\partial_\alpha \hat{\bf q}_j(x,\alpha)| \{I_i\}_{i=1}^N\rangle=\frac{1}{L} \langle \{I_i\}_{i=1}^N|\partial_\alpha \hat{\mathcal{Q}}_j(\alpha)| \{I_i\}_{i=1}^N\rangle\, .
\ee
The expectation value $\langle \{I_i\}_{i=1}^N|\partial_\alpha \hat{\mathcal{Q}}_j(\alpha)| \{I_i\}_{i=1}^N\rangle$ can be computed using the fact that $| \{I_i\}_{i=1}^N\rangle$ is an eigenstate of the charge
\be
\hat{\mathcal{Q}}_j(\alpha)| \{I_i\}_{i=1}^N\rangle=\left(\sum_{i=1}^N q_j(\lambda_i,\alpha)\right)| \{I_i\}_{i=1}^N\rangle\,
\ee
and generalizing the Hellman-Feynman theorem
\be
\partial_\alpha\left(\langle \{I_i\}_{i=1}^N|\hat{\mathcal{Q}}_j(\alpha)| \{I_i\}_{i=1}^N\rangle\right)=\left(\sum_{i=1}^N q_j(\lambda_i,\alpha)\right)\partial_\alpha\Big(\langle  \{I_i\}_{i=1}^N| \{I_i\}_{i=1}^N\rangle\Big)+\langle \{I_i\}_{i=1}^N|\partial_\alpha \hat{\mathcal{Q}}_j(\alpha)| \{I_i\}_{i=1}^N\rangle\, .
\ee
Above, the derivative is taken keeping the Bethe integers fixed.
Since the norm of the state is constant, we get the identity $\partial_\alpha\left(\langle \{I_i\}_{i=1}^N|\hat{\mathcal{Q}}_j(\alpha)| \{I_i\}_{i=1}^N\rangle\right)=\langle \{I_i\}_{i=1}^N|\partial_\alpha \hat{\mathcal{Q}}_j(\alpha)| \{I_i\}_{i=1}^N\rangle$.
Taking the derivative of the expectation value of the charge, we get two effects: one due to the parametric change of the charge eigenvalues, the other due to a rearrangement of the rapidities caused by a modification of the scattering phase shift in Eq. \eqref{S_eq_BetheGaudin}
\be\label{S_hf_notyet}
\partial_\alpha\left(\langle \{I_i\}_{i=1}^N|\hat{\mathcal{Q}}_j(\alpha)| \{I_i\}_{i=1}^N\rangle\right)=\partial_\alpha\left(\sum_{i=1}^N q_j(\lambda_i,\alpha)\right)=\sum_{i=1}^N \partial_\alpha q_j(\lambda_i,\alpha)+\partial_\alpha\lambda_i\partial_\lambda q_j(\lambda_i,\alpha)\, .
\ee
When the thermodynamic limit is enforced, the first term above simply becomes
\be
\lim_{\text{TDL}} \sum_{i=1}^N \partial_\alpha q_j(\lambda_i,\alpha)=L\int \dd \lambda\, \partial_\alpha q_j(\lambda,\alpha) \rho(\lambda)\, .
\ee
Instead, the second term requires extra manipulations. Indeed, deriving the Bethe Gaudin equations \eqref{S_eq_BetheGaudin} we get

\be
\partial_\alpha \lambda_i\left[\partial_\lambda p(\lambda_i)-\frac{1}{L}\sum_{j\ne i} \partial_{\lambda_i}\Theta(\lambda_i-\lambda_j)\right]=-\partial_\alpha p(\lambda_i)+\frac{1}{L}\sum_{j\ne i} \partial_\alpha\Theta(\lambda_i-\lambda_j)
-\frac{1}{ L}\sum_{j\ne i} \partial_{\lambda_i}\Theta(\lambda_i-\lambda_j)\partial_\alpha \lambda_j\,
\ee

In the thermodynamic limit, the above equation becomes
\be
\partial_\alpha \lambda_i = \frac{f^\text{dr}(\lambda_i)}{(\partial_\lambda p(\lambda_i))^\text{dr}}\, ,
\ee
with
\be
f(\lambda)=-\partial_\alpha p(\lambda)+\int \frac{\dd\mu}{2\pi}\partial_\alpha\Theta(\lambda-\mu)\vartheta(\mu)(\partial_\mu p)^\text{dr}\, ,
\ee
i.e. Eq. (5)  presented in the main text.
Replacing the last finding  into Eq. \eqref{S_hf_notyet} we finally get
\be
\langle \partial_\alpha \hat{\bf q}_j(x,\alpha)\rangle=\frac{1}{L}\partial_\alpha\left(\langle \{I_i\}_{i=1}^N|\hat{\mathcal{Q}}_j(\alpha)| \{I_i\}_{i=1}^N\rangle\right)=\int \dd\lambda\, \partial_\alpha q_j(\lambda,\alpha)\rho_j(\lambda)+ \frac{1}{2\pi}f^\text{dr}(\lambda) \partial_\lambda q_j(\lambda,\alpha)\vartheta(\lambda)\,,
\ee
i.e. Eq. \eqref{S_eq_HF}, as we desired.

\section{Numerical solution of the GHD equation}
\label{SM_sec_num}

This section is dedicated to numerical methods for solving the GHD equation. It is convenient to look at the equation in terms of the filling \eqref{S_eq_GHD_2}.
Interestingly, it admits the following implicit solution
\be\label{S_eq_implicit_solution}
\vartheta(t',x,\lambda)=\vartheta(t,x(t',t),\lambda(t',t))\,
\ee
where
\be\label{S_eq_traj}
x(t',t)=x-\int_t^{t'}\dd \tau\, v^\text{eff}_\tau(x(\tau,t),\lambda(\tau,t))\, \hspace{2pc} \lambda(t',t)=\lambda-\int_t^{t'}\dd \tau\, \left[\frac{\partial_\tau \alpha f^\text{dr}+\partial_x\alpha \Lambda^\text{dr}}{(\partial_\lambda p)^\text{dr})}\right]_{(\tau,x(\tau,t),\lambda(\tau,t))}\, .
\ee
Above, the effective velocity and the forces must be computed at the integration time, i.e. using the root density and the coupling at that time. Furthermore, they must be computed along the trajectories $(x(\tau,t),\lambda(\tau,t))$. Checking that Eq. \eqref{S_eq_implicit_solution} satisfies the GHD equation is immediate, however the solution is only implicit, since Eq. \eqref{S_eq_traj} depends on $\vartheta$ through the dressing operations. Nevertheless, the implicit solution is very useful in constructing numerical algorithms. We introduce a time step $\dd t$ and we are interested in updating the filling from $t$ to $t\to t+ \dd t$, thus we write
\be\label{S_micro_update}
\vartheta(t+\dd t,x,\lambda)=\vartheta(t,x(t+\dd t,t),\lambda(t+\dd t,t))\, .
\ee
Eq. \eqref{S_micro_update} is in principle exact for any $\dd t$: errors are introduced only when we approximate $x(t+\dd t,t)$ and $\lambda(t+\dd t,t)$.

\subsection{A first order method}
A first order method is readily obtained with the crude approximation
\begin{eqnarray}
x(t+\dd t,t)&=&x-\dd t\, v^\text{eff}_t(x,\lambda)+O(\dd t^2)\,, \\ \nonumber &&\\ \lambda(t+\dd t,t)&=&\lambda-\dd t\, \left[\frac{\partial_\tau \alpha f^\text{dr}+\partial_x\alpha \Lambda^\text{dr}}{(\partial_\lambda p)^\text{dr})}\right]_{(t,x,\lambda)}+O(\dd t^2)\, .
\end{eqnarray}

This provides a $O(\dd t)$ algorithm: indeed, at any update of the fillings an error $O(\dd t^2)$ is introduced and, in order to reach a time $t$, we need $t/\dd t$ steps. Therefore, at time $t$ we accumulate an error $\sim t \dd t$.
This method has already been proposed in Ref. \cite{S_GHD7} (albeit in absence of force terms). In Ref. \cite{S_GHD7} it has been observed that estimating the integrals appearing in $x(t+\dd t,t)$ $\lambda(t+\dd t,t)$ with the endpoints of the leap (instead of the starting one as per above) makes the algorithm more stable, which however remains first order in time.
Here, we further improve the algorithm providing a $O(\dd t^2)$ method.

\subsection{A second order method}

A better approximation for $x(t+\dd t, t)$ and $\lambda(t+\dd t, t)$ can be obtained taking the middle points in the integrals \eqref{S_eq_traj} rather than the extrema. Therefore

\begin{eqnarray}\label{S-eq_shift1}
x(t+\dd t,t)&=&x-\dd t\, v^\text{eff}_{t+\dd t/2}(x',\lambda')+O(\dd t^3)\,, \\ \nonumber &&\\
\label{S-eq_shift2} \lambda(t+\dd t,t)&=&\lambda-\dd t\, \left[\frac{\partial_\tau \alpha f^\text{dr}+\partial_x \alpha \Lambda^\text{dr}}{(\partial_\lambda p)^\text{dr})}\right]_{(t+\dd t/2,x',\lambda')}+O(\dd t^3)\, ,
\end{eqnarray}
with $x'=x(t+\dd t/2,t)$ and $\lambda'=\lambda(t+\dd t/2,t)$. The exact expressions for $x'$ and $\lambda'$ are unknown, but we can estimate them at first order

\begin{eqnarray}
x'&=&x-\frac{\dd t}{2}\, v^\text{eff}_{t+\dd t/2}(x,\lambda)+O(\dd t^2)\,, \\ \nonumber &&\\ \lambda'&=&\lambda-\frac{\dd t}{2}\, \left[\frac{\partial_\tau \alpha f^\text{dr}+\partial_x\alpha \Lambda^\text{dr}}{(\partial_\lambda p)^\text{dr})}\right]_{(t+\dd t /2,x,\lambda)}+O(\dd t^2)\, .
\end{eqnarray}
This approximation contributes with a $O(\dd t^3)$ correction to (\ref{S-eq_shift1}-\ref{S-eq_shift2}).
Using middle points in the time leap requires computing the fillings at times $n \dd t$ and $(n+1/2) \dd t$, where $n$ is an integer. Therefore, this method is two time slower than the first order one if the same time step $\dd t$ is used, but the global error grows as $\sim t \dd t^2$. So, in general cases this algorithm outclasses the previous one, since much larger time steps can be considered.

There is a subtlety in this algorithm that needs to be mentioned, i.e. we need the fillings at times $t=0$ and $t=\dd t/2$ as a starting points. While the filling at time $t=0$ is simply the initial condition and provided by the TBA solution, the filling at $t=\dd t/2$ is not. In order to determine it, we choose a second time step $\dd t'\ll \dd t/2$ and approximate $\vartheta(t+\dd t'/2,x,\lambda)$ according to the first order algorithm, then the filling is evolved with the second order algorithm up to time $t+\dd t/2$ using a time step $\dd t'$. At this point, both $\vartheta(t,x,\lambda)$ and $\vartheta(t+\dd t/2,x,\lambda)$ are known and we can proceed with the second order algorithm with time step $\dd t$.

\section{The TBA of the models of interest}
\label{SM_sec_TBA}

In this short Section, for the sake of completeness, we briefly review the TBA description of the models we looked at. For a more detailed presentation of the TBA method, the reader can refer to Ref. \cite{S_taka}.
For any model, we also shortly mention the details of the numerical methods used in making the plots presented in the main text.

\subsection{The interacting Bose igas}

The interacting Bose gas describes bosons with contact interaction and it is known to be integrable since a long time \cite{S_LiLi63,S_LiLi63_bis}. Within the second quantization formalism, its Hamiltonian reads

\be
\hat{H}=\int_{0}^{L}\,{\rm d}x \left\{\frac{1}{2m}\partial_x\hat{\psi}^{\dagger}(x)\partial_x\hat{\psi}(x)+c\hat{\psi}^{\dagger}(x)\hat{\psi}^{\dagger}(x)\hat{\psi}(x)\hat{\psi}(x)-\mu \hat{\psi}^\dagger(x)\hat{\psi}(x)\right\}\,,
\label{eq:hamiltonian}
\ee
The fields $\hat{\psi}^{\dagger}(x)$,$\hat{\psi}(x)$ are bosonic creation and annihilation operators $\left[\hat{\psi}(x),\hat{\psi}^{\dagger}(y)\right]=\delta(x-y)$. The interaction strength is assumed to be positive $c>0$ and we explicitly introduced the chemical potential $\mu$ which, once it is made inhomogeneous, can describe external traps.

Within the repulsive regime, the model does not have bound states, therefore its TBA is formulated in terms of a single species of particle with bare energy and momentum given by
\be
\epsilon(\lambda)=\frac{\lambda^2}{2m}-\mu\,, \hspace{3pc}p(\lambda)=\lambda\, .
\ee
The rapidity lives on the whole real line $\lambda \in (-\infty,\infty)$.
The expectation value of energy and density of particles, which are the observables on which we focus on, are (the thermodynamic limit is always enforced)
\be
\frac{1}{L}\langle \hat H\rangle=\int\dd\lambda\, \epsilon(\lambda)\rho(\lambda)\,,\hspace{2pc}\langle \hat \psi^\dagger(x)\hat \psi(x)\rangle=\int \dd \lambda \, \rho(\lambda)\, .
\ee

Analyzing the model by mean of coordinate Bethe Ansatz, the following scattering matrix can be derived
\be\label{eq_LL_S}
S_\text{LL}(\lambda)=\frac{\lambda+2im c}{\lambda-2im c}\hspace{2pc}\Longrightarrow\hspace{2pc} \Theta(\lambda)=\arctan\left(\frac{4\lambda m c}{\lambda^2-(2m c)^2}\right)\, .
\ee
Thermal states can be described according to Eq. \eqref{eq:filling} and Eq. \eqref{eq:thermal_TBA}.

\subsubsection{Details of the numerical simulations}

In Fig. 2 we numerically simulated an interaction quench for a trapped interacting Bose gas. The GHD equations are solved with the second order method presented in Section \ref{SM_sec_num} using a time step $\dd t=0.025$.
The instantaneous TBA equations are solved by discretizing the integrals using Gaussian quadratures, thus converting the linear integral equations into finite-dimensional vector-matrix equations. The rapidity space has a cut off $|\lambda|\le 3$ and its discretized on a lattice of $100$ points. The spatial coordinates are taken within the interval $x\in[-3,3]$ and are discretized on a lattice of $100$ points, with constant lattice space.
In order to check the precision of the solution, we monitored the conservation of the total number of particles which is constant 
with $\lesssim 0.5\% $ fluctuations over the explored time scales.

\subsection{The XXZ spin chain}

The XXZ spin chain is governed by the Hamiltonian
\be
\hat H= \sum_{j=1}^N \{\hat S_j^x \hat S_{j+1}^x +\hat S_j^y \hat S_{j+1}^y  + \Delta \hat S_j^z \hat S_{j+1}^z  +B \hat S_j^z \}\,.
\ee
Above, $\hat S_j^{x,y,z}$ are usual spin$-\frac{1}{2}$ operators.
Differently from the Lieb Liniger model, the XXZ spin chain always supports bound states and thus the TBA requires multiple root densities. The thermodynamics is greatly affected by the value of $\Delta$, in particular the cases $|\Delta|<1$ and $|\Delta|\ge 1$ require a different discussion.
For $|\Delta|<1$ the TBA has a fractal dependence on the value of $\Delta$ \cite{S_taka}. For this reason, inhomogeneous space-time dependent $\Delta-$profiles within this phase lay outside of the applicability of our method, which requires a smooth dependence of the model on the coupling.

Instead, the $|\Delta|\ge 1$ case is not pathological: more specifically, we focus on $\Delta\ge 1$ and in the positive magnetization sector $B<0$ (which implies $\langle S^z_j\rangle >0$).
The TBA description requires infinitely many root densities, usually called strings, $\{\rho_j(\lambda)\}_{j=1}^\infty$.
Accounting for several strings in the TBA is straightforward.

In the $\Delta\ge 1$ case the rapidities are confined to a Brillouin zone $\lambda\in [-\pi/2,\pi/2]$. To each string are associated an energy $\epsilon_j(\lambda)$ and a momentum $p_j(\lambda)$ ($j\in \{1,2,3,...\}$)
\be
\epsilon_j(\lambda)=-\frac{1}{2} \sin(\theta) \partial_\lambda p_j(\lambda)- j B\, , \hspace{2pc}p_j(\lambda)=2 \arctan\left[\coth\left(\frac{j \theta}{2}\right)\tan \lambda\right]\, ,
\ee

where the angle $\theta$ parametrizes the coupling $\Delta=\cosh\theta$.
Among the possible relevant observables, the expectation values of the Hamiltonian and local magnetization are of outmost simplicity
\be
\frac{1}{N}\langle \hat H\rangle=\frac{\Delta}{4}+\sum_j \int_{-\pi/2}^{\pi/2} \dd \lambda \epsilon_j(\lambda) \rho_j(\lambda)\, , \hspace{2pc}\langle \hat S_i^z\rangle=\frac{1}{2}-\sum_j\int_{-\pi/2}^{\pi/2} \dd \lambda\, j\rho_j(\lambda)\, .
\ee

The scattering phase is promoted to be a matrix with indexes running over all the possible strings
\be
\Theta_{j,k}(\lambda)=(1-\delta_{j,k})\frac{p_{|j-k|}(\lambda)}{2\pi}+\frac{p_{j+k}(\lambda)}{2\pi}+2\sum_{\ell=1}^{\min(j,k)-{\red 1}}\frac{p_{|j-k|+2\ell}(\lambda)}{2\pi}\, .
\ee
The dressing operation now keeps in account the presence of multiple strings, therefore a function $\tau_j(\lambda)$ is now dressed according to
\be
\tau_j^\text{dr}(\lambda)=\tau_j(\lambda)-\sum_{i}\int_{-\pi/2}^{\pi/2} \frac{\dd \mu}{2\pi} \partial_\lambda \Theta_{ji}(\lambda-\mu)\vartheta_i(\mu)\tau^\text{dr}_i(\mu)\, .
\ee

The thermal states are now described by the set of equations
\be
 \vartheta_j(\lambda)=\frac{1}{e^{\varepsilon_j(\lambda)}+1}\,,
\ee
\be
\varepsilon_j(\lambda)=\beta\epsilon_j(\lambda)+\sum_i\int_{-\pi/2}^{\pi/2}\frac{{\rm d}\mu}{2\pi}\partial_\lambda\Theta_{j,i}(\lambda-\mu)\log\left(1+e^{-\varepsilon_i(\mu)}\right)\,.
\ee

Notice that for $B<0$ the fillings are exponentially vanishing while increasing the string index $j$, thus the infinite set of strings can be truncated only to the first ones, the quality of the approximation being decided by the magnetic field $B$ and the inverse temperature $\beta$.
We mention that the ground state, i.e. $\beta\to\infty$, is such that $\vartheta_{j>2}(\lambda)=0$, thus we can use only the first string to describe it.

\subsubsection{Details of the numerical simulations}

In Fig. 3 we provide a benchmark of the GHD equations against tDMRG~\cite{S_itensor} simulations.
For what it concerns the GHD simulations, with the parameters we choose (i.e. low temperature) we 
found that retaining only the first two strings gives a satisfactory precision. The rapidity 
space is discretized into $50$ points and integral equations are solved by means of Gauss 
quadratures. The position lives on an interval $[-1,1]$ which is discretized into $100$ 
equally spaced lattice points. The time evolution is solved  according to the second 
order algorithm with time step $\dd t=0.0125$.

For the tDMRG simulations we employed the standard purification method~\cite{S_Sch05} to represent the initial 
density matrix. The time evolution was implemented by using the MPO representation for the evolution 
operators $e^{- i Ht}$. To mitigate the error associated with the time discretization we employed the 
scheme presented in Ref.~\cite{S_greg}, which allows one to obtaine an accuracy ${\mathcal O}(dt^5)$. 
The application of the MPO evolution operator is implemented by using the fitting algorithm described 
in Ref.\cite{S_metts}. In our simulations we used $dt=0.1$. The maximum bond dimension employed was 
$\chi\approx 500$.

\subsection{The classical sinh-Gordon model}

The sinh-Gordon model is a relativistic field theory of a scalar field $\phi$, governed by the Lagrangian
\be
\mathcal{L}=\int \dd x \, \frac{1}{2}\partial_\mu\phi\partial^\mu\phi-\frac{m^2}{g^2}\big[\cosh (g\phi)-1\big]\, .
\ee
The model is integrable both at classical and quantum level. The reader could be more familiar with the thermodynamics of the quantum system, but a proper semiclassical limit of the latter readily gives access to the GGE \cite{S_DeLucaMuss16} and GHD \cite{S_BDWY17} of the classical version.
We leave to the original references the details and present here the relevant results.

The classical shG model can be described in terms of a single species of particle, having energy and momentum eigenvalues given by
\be
\epsilon(\lambda)=m\cosh\lambda\,, \hspace{2pc} p(\lambda)=m\sinh\lambda\, .
\ee
Notice that, in contrast with the quantum case, no renormalization of the mass occurs and the single particle eigenvalues are independent from the interaction $g$.
The scattering phase $\Theta$ is singular and defined as
\be
\Theta_\gamma(\lambda) =\frac{g^2}{8}\left[\frac{1}{\sinh\lambda+i \gamma}+\frac{1}{\sinh\lambda-i \gamma}\right]\, ,
\ee
where the limit $\gamma\to 0^+$ must be enforced after the integrations have been carried out. For example, the dressing operation is actually defined as
\be
\tau^\text{dr}(\lambda)=\tau(\lambda)-\lim_{\gamma\to 0^+}\int \frac{\dd \lambda}{2\pi}\partial_\lambda\Theta_\gamma(\lambda-\mu)\vartheta(\mu)\tau^\text{dr}(\mu)\, .
\ee
The filling of thermal states is written in terms of the effective energy $\varepsilon(\lambda)$ as
\be
\vartheta(\lambda)=\frac{1}{\varepsilon(\lambda)}\, ,
\ee
which satisfies the following integral equation
\be
\varepsilon(\lambda)=\beta\epsilon(\lambda)-\lim_{\gamma\to 0^+}\int \frac{\dd \mu}{2\pi} \partial_\lambda \Theta_\gamma(\lambda-\mu) \log \varepsilon(\mu)\, .
\ee

The expectation value of the energy is UV divergent on thermal states, similarly to what it happens in the famous black-body catastrophe. For this reason, we revert to other local operators with well-defined UV properties, namely the vertex operators $e^{k g\phi}$. Their expectation values on arbitrary GGEs are recursively fixed by the following set of integral equations \cite{S_BDWY17}
\be
\frac{\langle e^{(k+1)g\Phi}\rangle}{\langle e^{kg\Phi}\rangle}=1+(2k+1)\frac{g^2}{4\pi}\int \dd\lambda\,\, e^\lambda \vartheta(\lambda) \xi^k(\lambda)\, \, ,\label{clvertex}
\ee
where
\be
\xi^k(\lambda)=e^{-\lambda}+\frac{ g^2}{4} \mathcal{P}\int \frac{\dd\mu}{2\pi} \frac{1}{\sinh(\lambda-\mu)}\left(2k-\partial_\mu \right)(\vartheta(\mu)\xi^k(\mu))\, \, .\label{clpint}
\ee
Above $\mathcal{P}$ stands for the principal value regularization of the singular integral. Eq. \eqref{clvertex} allows for a recursive determination of $\langle e^{k g\phi}\rangle$ for $k=1,2,...$ using the fact that, for $k=0$, the vertex operator becomes the identity $\langle e^{kg\phi}\rangle\Big|_{k=0}=\langle 1\rangle=1$.

\subsubsection{Details of the numerical simulations}

In Fig. 4 we compare the GHD predictions against Monte Carlo simulations. The GHD is solved with the second order algorithm of Section \ref{SM_sec_num} with time step $\dd t =0.025$. The singular nature of the integral equations requires a careful discretization whose details can be found in Ref. \cite{S_BDWY17}: for our purposes, we restricted the rapidities on a finite interval $[-10,10]$ which is discretized into $200$ equispaced lattice points. The spatial direction is restricted on the interval $[-1,1]$ which is discretized on a lattice of $200$ equispaced points.

The shG model is directly simulated through Metropolis-Hasting techniques presented in Ref. \cite{S_BDWY17}: the interval $[-L,L]$, together with the temporal direction, is discretized on a tilted squared lattice (lattice space $a=0.025$ and length $L=30$, i.e. $600$ points in the spatial direction). The initial configurations are sampled from a (inhomogeneous) thermal ensemble generated through a Metropolis-Hasting algorithm. Subsequently, each initial configuration is then deterministically evolved in time: observables are then averaged on the initial conditions. We took roughly $3.5\times 10^5$ realizations.

\section{The interacting Bose gas as non relativistic limit of the sinh-Gordon model}
\label{suppl_NR_sec}

The derivation of the GHD equations can be transferred, without the need of any ansatz, from the relativistic world to the non relativistic one, through proper non relativistic limits. In this short section, for the sake of completeness, we mention how this operation can be performed on the interacting Bose gas, viewed as the non relativistic limit of the quantum sinh-Gordon model.
We have already briefly presented the interacting Bose gas and the classical sinh-Gordon model in the previous section, here we must now revert to the quantum sinh-Gordon model explicitly restoring the speed of light $c_\text{light}$, which will be then send to infinity. The interacting Bose gas has been identified as the NR limit of the shG model in Ref. \cite{S_KoMuTr10,S_KoMuTr09}, then the same approach has been extended to a larger class of models in Ref. \cite{S_BaMuDeLu16,S_BaMuDeLu17}. Here, we leave to the original references a careful treatment of the limit, presenting only the most important steps.
The Lagrangian of the model is
\be
\mathcal{L}_\text{shG}=\int \dd x\,  \frac{1}{2c_\text{light}^2}(\partial_t\phi)^2-\frac{1}{2}(\partial_x\phi)^2-\frac{m^2c_\text{light}^4}{16 c }(\cosh(c_\text{light}^{-1}4\sqrt{c}\phi)-1)\, .
\label{shGaction}
\ee
Above, we restored the speed of light and redefine the interaction $g$ in such a way to make a direct contact with the interacting Bose Gas. The identification is achieved in the limit $c_\text{light}\to \infty$ that, in terms of the rescaled interaction in the shG model, corresponds also to the weakly interacting limit.
The identification is readily seen at the level of scattering matrix. Indeed, the shG quantum model possesses an unique species of excitation with scattering matrix
\be
S_{\text{shG}}(\theta)=\frac{\sinh\theta-i\,\sin(\pi\alpha)}{\sinh\theta+i\,\sin(\pi\alpha)}\,\,\,,
\ee
where the parameter $\alpha$ is
\be
\alpha \,= \,\frac{c_\text{light}^{-1}16 c}{8\pi + c_\text{light}^{-1}16 c} \,\,\,.
\ee 

Since the relativistic momentum is $p(\theta)=c_\text{light}M \sinh \theta$ (with $M$ the renormalized mass), posing $\theta\sim \lambda /(mc_\text{light})$ (where we use that in the weakly interacting limit the renormalized mass tends to the bare one and that the rapidity in the interacting Bose gas is simply the momentum) we readily find that the shG scattering matrix collapses to the interacting Bose gas's one
\be
\lim_{c_\text{light}\to \infty}S_{\text{shG}}(\lambda/(mc_\text{light}))=S_\text{LL}(\lambda)\,,
\ee
with the r.h.s. being the scattering matrix of the interacting Bose gas Eq. \eqref{eq_LL_S}.
The mapping can be extended to the whole Thermodynamic Bethe Ansatz \cite{S_KoMuTr10,S_KoMuTr09}, where it has been understood that the filling of the shG model simply becomes that of the interacting Bose gas, once the NR limit has been taken
\be\label{NR_fill}
\lim_{c_\text{light}\to\infty}\vartheta_{\text{shG}}(\lambda/(mc_\text{light}))=\vartheta_\text{LL}(\lambda)\, .
\ee
This makes very easy to consider the non relativistic limit of dressed quantities. For example, let us consider a test function $\tau_\text{shG}(\theta)$ in the shG model and the relative dressing
\be
\tau_\text{shG}^\text{dr (shG)}(\theta)=\tau_\text{shG}(\theta)-\int \frac{\dd \mu}{2\pi} \partial_\theta \Theta_\text{shG}(\theta-\mu)\vartheta_\text{shG}(\mu)\tau_\text{shG}^\text{dr (shG)}(\mu)\, ,
\ee
Replacing $\theta=\lambda/(mc_\text{light})$ and using Eq. \eqref{NR_fill}, we simply have, for example, for the derivative of the momentum
\be
\lim_{c_\text{light}\to\infty}\left[\frac{1}{mc_\text{light}}(\partial_\theta p_\text{shG})^\text{dr (shG)}(\lambda/(mc_\text{ligth}))\right]= (\partial_\lambda p_\text{LL})^\text{dr (LL)}(\lambda)
\ee
where on the l.h.s we consider the momentum on the shG model dressed according to the shG TBA, while on the r.h.s we have the interacting Bose gas momentum and the dressing operation according to that model.
It is now a matter of a simple exercise to extend the above identity to all the terms appearing in the GHD equation, obtaining the hydrodynamic of the interacting Bose gas as the non relativistic limit of that of the quantum sinh-Gordon model.

\end{document}